\documentclass[aps,prb,preprint,groupedaddress,nobibnotes]{revtex4-2}

\usepackage{amsmath,amsfonts,amssymb}
\usepackage{graphicx}
\usepackage{siunitx}
\usepackage{glossaries}
\usepackage{microtype}
\usepackage[colorlinks,allcolors={blue}]{hyperref}
\usepackage[english]{babel}
\usepackage[utf8]{inputenc}
\usepackage[sort&compress]{natbib}

\newacronym{cv}{CV}{cross validation}
\newacronym{qnm}{QNM}{quasinormal mode}
\newacronym{nn}{NN}{neural network}
\newacronym{phc}{PhC}{photonic crystal}
\newacronym{ms}{MS}{metasurface}
\newacronym{sm}{SI}{Supporting Information}

\begin{document}

\title{A General Framework for Knowledge Integration in Machine Learning for Electromagnetic Scattering Using Quasinormal Modes}

\author{Viktor A. Lilja}
\author{Albin J. Svärdsby}
\author{Timo Gahlmann}
\author{Philippe Tassin}

\affiliation{Department of Physics, Chalmers University of Technology, SE-41296 Göteborg, Sweden}

\date{\today}

\maketitle

\section*{Abstract}
Neural networks have been demonstrated to be able to accelerate the modeling and inverse design of optical and electromagnetic devices by serving as fast surrogates for electromagnetic solvers. 
Nevertheless, such neural networks can be unreliable and normally require extreme amounts of data to train. 
Here it is shown that these limitations can be alleviated by constraining neural-network models using prior knowledge about the governing physics. We propose a universal physics-informed neural network framework for electromagnetic scattering based on the quasinormal mode expansion of the scattering matrix.
The neural networks learn the resonant structure underlying the scattering spectrum, are guaranteed to obey energy conservation and causality, and are shown to have significantly improved data efficiency for photonic-crystal slabs and all-dielectric free-form metasurfaces.
Furthermore, the framework allows additional problem-specific constraints, such as losslessness, symmetries, and number of modes, to be imposed manually when they are available.
The method can be applied to a wide range of optical and electromagnetic devices owing to the generality of the quasinormal mode formalism.

\section{Introduction}
\label{sect:intro}
Machine learning~\cite{daSilva_book_2017, mehlig_ML_2021} is a powerful tool for inverse design in electromagnetism, mechanics, materials science, quantum many-body systems, and other paradigms of physics~\cite{wiecha_storage_2019,iten_discoveringphysics_2020, peano_topological_2021, wiecha_deep_2021, khatib_deep_2021, jiang_deep_2021, zandehshahvar_reduction_2021, zandehshahvar_manifold_2022, mohseni_quantummanybody_2022, lee_2023, sanchez_advances_2024, xu_physics-informed_2024}.
The goal of inverse design is to find a device design that exhibits some predefined properties~\cite{su_nanophotonic_2020, tsilipakos-two-photon_2024, hegde_inverse_2019, so_deep_2020, deng_neural-adjoint_2021}.
For linear electromagnetic devices, the properties are often quantities that describe the relation between incoming and outgoing amplitudes of electromagnetic modes, such as transmittance, reflectance, or absorption~\cite{wiecha_deep_2021, khatib_deep_2021, jiang_deep_2021, xu_physics-informed_2024, so_deep_2020, deng_neural-adjoint_2021, yan_highly_2024, nadell_deep_2019, sarkar_physics-informed_2023, xu_enhanced_2020, jing_deep_2023, blanchard-dionne_teaching_2020, zhang_symmetry_2022, so_simultaneous_2019, ma_strategical_2022, khatib_learning_2022, ma_probabilistic_2019, guimbao_numerical_2022, liu_generative_2018, xu_freeform_2024, gahlmann_deep_2022, yeung_deepadjoint_2023, tahersima_deep_2019, qian_deep_2023, liu_training_2018, kildishev_chiral_2024, you_driving_2024, gao_meta-attention_2024}.
These quantities can be derived from the frequency-dependent scattering matrix $S(\omega)$, which encodes the complete scattering response of a general linear $N$-port device by relating the amplitudes $\mathbf{s}^- = (s_1^-,\ldots,s_N^-)$ of the incoming modes to the amplitudes $\mathbf{s}^+ = (s_1^+,\ldots,s_N^+)$ of the outgoing modes through $\mathbf{s}^- = S(\omega) \mathbf{s}^+$.
The forward problem of finding $S(\omega)$ for a given design can be easily solved with full-wave electrodynamics solvers.
The inverse problem of finding a design with a desired $S(\omega)$ is, on the other hand, significantly more challenging and fundamentally ill-posed, since a given $S(\omega)$ can have zero or several corresponding designs. Traditional inverse-design methods use optimization algorithms to search through the design space, often requiring hundreds to thousands of sequential full-wave simulations to converge.
Another approach is to first replace the simulator with a \gls{nn} trained to predict $S(\omega)$ given a design parametrization, summarized schematically as
\begin{equation*}
    \textnormal{Design} 
    \xrightarrow[]{\textnormal{NN}}
    S(\omega).
\end{equation*}
Once trained, this forward model can typically predict $S(\omega)$ orders of magnitude faster than an electrodynamics solver and its gradient can easily be calculated using backpropagation.
This allows for efficient exploration of the design space through iterative optimization~\cite{deng_neural-adjoint_2021, su_metaphynet_2024} or the training of inverse models~\cite{sarkar_physics-informed_2023, ma_probabilistic_2019, liu_generative_2018, xu_freeform_2024, gahlmann_deep_2022, liu_training_2018, kudyshev_global_2021}.
The downside of this approach is that the performance of the final design becomes dependent on the accuracy of the \gls{nn}.
Although it may be possible to improve accuracy by increasing the number of trainable parameters and the number of training samples, computational resources are often a limiting factor~\cite{sanchez_advances_2024}.
Furthermore, the black box nature of \glspl{nn} makes them unreliable and very limited insights can be gained from their predictions~\cite{kim_knowledge_2021}.

These issues can be alleviated by utilizing prior knowledge about the governing physics~\cite{karniadakis_physics-informed_2021, kim_knowledge_2021}, a strategy often referred to as physics-informed machine learning.
One physics-informed method, which has recently gathered attention in electromagnetic scattering problems~\cite{xu_physics-informed_2024, yan_highly_2024, blanchard-dionne_teaching_2020, khatib_learning_2022}, is to include a physics layer as a final step in the forward model. 
The augmented forward model can be schematically summarized as
\begin{equation}\label{eq:physics-informed-principle}
    \textnormal{Design} 
    \xrightarrow[]{\textnormal{NN}}
    \begin{matrix}
        \textnormal{Physics} \\
        \textnormal{parameters}
    \end{matrix}
    \xrightarrow[]{\textnormal{Physics model}}
    S(\omega),
\end{equation}
where the physics model is a known computable expression for $S(\omega)$ with the physics parameters as arguments.
Now, any constraints imposed on the physics model will also be satisfied by the full forward model.
Furthermore, the task of the \gls{nn} has become to predict the physics parameters rather than $S(\omega)$ directly, which may be easier to learn.
Indeed, previous works have demonstrated that using a physics layer reduces prediction error for a given \gls{nn} size~\cite{xu_physics-informed_2024, blanchard-dionne_teaching_2020, khatib_learning_2022}, improves generalization~\cite{yan_highly_2024, blanchard-dionne_teaching_2020, khatib_learning_2022}, and reduces the required amount of training data~\cite{yan_highly_2024, khatib_learning_2022}.
Another benefit is that the physics parameters become accessible as a learned latent description~\cite{yan_highly_2024, khatib_learning_2022} that can be used as a design target in the inverse problem~\cite{xu_physics-informed_2024}.
The forward model also becomes less of a black box in the sense that the predicted spectra can be explained in terms of the physics parameters.
Implementing the physics model in a numerical framework that supports automatic differentiation enables the full forward model to be trained directly on simulated $S(\omega)$.
This is in contrast to the methods proposed in References~\cite{ma_strategical_2022, su_metaphynet_2024, zhang_physics-driven_2023, ding_neural-network_2004, liu_novel_2024}, where the physics parameters are extracted from $S(\omega)$ before training.

Various physics models that exploit the resonant structure of the scattering matrix have been proposed in the literature~\cite{xu_physics-informed_2024, blanchard-dionne_teaching_2020, khatib_learning_2022}, but are either tailored to specific device geometries~\cite{xu_physics-informed_2024, khatib_learning_2022}, making their range of applicability limited,  or rely on spectral decompositions without a formal theoretical basis~\cite{blanchard-dionne_teaching_2020}, making the physics parameters hard to interpret.
However, the general and exact relation between resonances and scattering parameters is, in fact, known and is formalized by the theory of \glspl{qnm}~\cite{alpeggiani_quasinormal-mode_2017, zhang_quasinormal_2020}.
We propose a physics-informed \gls{nn} architecture based on the \gls{qnm} expansion of the scattering matrix.
We will refer to this architecture as the QNM-Net.
Owing to the generality of the \gls{qnm} formalism, the QNM-Net is universal in the sense that it can be applied to linear electromagnetic devices with an arbitrary number of resonances and scattering ports.
Furthermore, by implementing the QNM-Net as a modular framework rather than a fixed architecture, we are able to use the rich physics of the \gls{qnm} formalism to manually tailor the model implementation to the unique physics of each problem. This makes our method simultaneously more general than previous methods while also allowing for stronger problem-specific constraints to be imposed.

\section{Quasinormal-mode-based Neural Network}
\label{sect:methods}
\glspl{qnm} are a generalization of normal modes to lossy systems.
They are defined as eigenmodes of the source-free Maxwell's equations 
\begin{equation}\label{eq:maxwell_equations}
\begin{split}
    \nabla\times\tilde{\mathbf{E}}_m &= i\tilde{\omega}_m\mu\tilde{\mathbf{H}}_m, \\
    \nabla\times\tilde{\mathbf{H}}_m &= -i\tilde{\omega}_m\varepsilon\tilde{\mathbf{E}}_m,
\end{split}    
\end{equation}
with outgoing radiation boundary conditions, where $\tilde{\omega}_m$ is the \gls{qnm} eigenfrequency, $m$ enumerates the \glspl{qnm}, and $\varepsilon$ and $\mu$ material parameters describing a linear and local material response.
Due to the open boundary and material absorption, the eigenproblem is non-Hermitian and $\tilde{\omega}_m = \omega_m + i \gamma_m$ complex-valued.
The imaginary part $\gamma_m$ corresponds to a decay in time and can be decomposed as $\gamma_m = \gamma_{\mathrm{r},m} + \gamma_{\mathrm{nr},m}$, where $\gamma_{\mathrm{r},m}$ is due to radiative losses and $\gamma_{\mathrm{nr},m}$ is due to nonradiative losses~\cite{lalanne_light_2018, kristensen_modeling_2020}.

\begin{figure}[hbt]
    \includegraphics{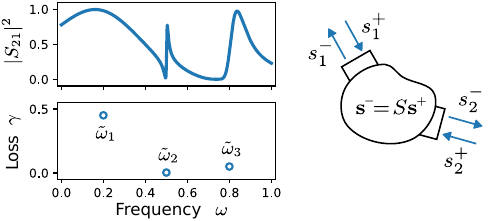}
    \caption{%
        Relation between the $S$ matrix and \gls{qnm}s of a fictitious 2-port scattering device. Right: The scattering matrix $S$ relates the amplitudes $s_n^\pm$ of incoming ($+$) and outgoing ($-$) waves. Top left: Resonances are observed when plotting elements of $S$ as a function of frequency. Bottom left: Each resonance corresponds to a complex pole of $S(\omega)$. The poles are eigenfrequencies $\tilde{\omega}_m$ of \glspl{qnm}.
    \label{fig:qnm-illustration}
    }
\end{figure}

A connection can be made between $S(\omega)$ and the \glspl{qnm} by expanding the scattered electromagnetic field as a sum of the \gls{qnm} fields $\tilde{\mathbf{E}}_m$ and $\tilde{\mathbf{H}}_m$~\cite{alpeggiani_quasinormal-mode_2017, zhang_quasinormal_2020, weiss_how_2018}.
This so-called quasinormal mode expansion of the scattering matrix makes it clear that the \gls{qnm} frequencies $\tilde{\omega}_m$ correspond to resonances in the scattering spectrum, as illustrated in \textbf{Figure~\ref{fig:qnm-illustration}}.
Many different versions of the expansion can be found in the literature~\cite{alpeggiani_quasinormal-mode_2017, weiss_how_2018, zhang_quasinormal_2020, benzaouia_quasi-normal_2021}.
In this work, we use the form 
\begin{equation}\label{eq:qnm-expansion}
    S(\omega) = e^{i\omega\tau}[C(\omega) + D(i\omega - i\tilde{\Omega})^{-1}M^{-1}D^\dagger C(\omega)]e^{i\omega\tau},
\end{equation}
which is approximate but specifically formulated to respect energy conservation even when a finite number of \glspl{qnm} are included in the expansion \cite{benzaouia_quasi-normal_2021}, making it suitable for our application.
In this expression, for an $N$-port device with $P$ modes, 
$C(\omega)$ is an $N \times N$ matrix representing a slowly varying background,
$D$ is an $N \times P$ matrix with columns $\mathbf{d}_m$, 
$\tilde{\Omega}$ is a diagonal matrix of \gls{qnm} frequencies $\tilde{\omega}_m$, and $M$ is a $P\times P$ matrix with elements
\begin{equation*}
    M_{mp} = \frac{\mathbf{d}_m^\dagger\mathbf{d}_p}{i\tilde{\omega}_{\mathrm{r},p} - i \tilde{\omega}_{\mathrm{r},m}^*},
\end{equation*}
where $\tilde{\omega}_{\mathrm{r},m} = \omega_m + i \gamma_{\mathrm{r},m}$.
The $N$ complex-valued elements of $\mathbf{d}_m$ should be interpreted as the amplitudes of the $m$:th \gls{qnm} on the ports.
Finally, $\tau$ is a diagonal $N \times N$ matrix of real numbers $\tau_n$ that determine the phase accumulated between the scatterer and the ports.
We refer to Reference~\cite{benzaouia_quasi-normal_2021} for a more in-depth description, derivation, and error analysis.

\begin{figure}[hbt]
    \includegraphics{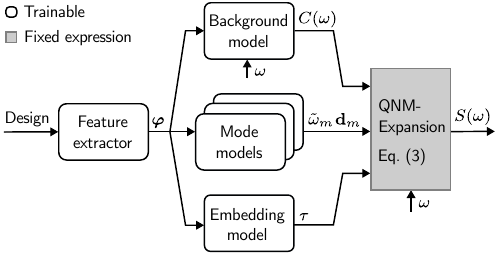}
    \caption{%
        Overview of the QNM-Net architecture. 
        White boxes are submodels with trainable parameters, the gray box is an analytical physics model with no trainable parameters.
        Arrows indicate the relations between inputs and outputs of each submodel.
        \label{fig:qnm-net-architecture}
    }
\end{figure}

Our proposed physics-informed \gls{nn} architecture incorporates the QNM expansion as a physics model according to the principle shown in the schematic in Equation~\eqref{eq:physics-informed-principle}.
The physics parameters are in our case $C(\omega)$ and $\tau$, as well as $\tilde{\omega}_m$ and $\mathbf{d}_m$ for each mode $m$.
We use a modular architecture composed of multiple submodels as illustrated in \textbf{Figure~\ref{fig:qnm-net-architecture}} to facilitate adaptation to a wide range of systems.
The first submodel maps the design description to an abstract feature vector $\boldsymbol{\varphi}$.
The subsequent submodels use $\boldsymbol{\varphi}$ to predict the physics parameters evaluated at frequency $\omega$.
Finally, $S(\omega)$ is computed from the physics parameters using the \gls{qnm} expansion~\eqref{eq:qnm-expansion}.
The internal structure of the submodels is left unspecified and can be tailored to the specific physics of each system.

\section{Results and Discussion}
\label{sec:results_and_discussion}

\subsection{Photonic crystal slab}
\begin{figure}[hbt!]
    \includegraphics{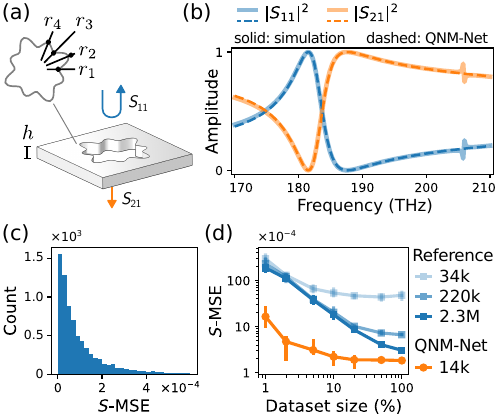}
    \caption{%
        Applying the QNM-Net to \gls{phc} slabs.
        (a)~Illustration of the \gls{phc} parametrization and geometry.
        (b)~Simulated and QNM-Net-predicted $S$ parameters for a sample from the validation set.
        (c)~Histogram of test losses for the QNM-Net when using \SI{20}{\percent} of the dataset.
        (d)~Validation loss as a function of dataset size for the QNM-Net and standard \glspl{nn}. Markers show cross-validation average, vertical bars show lowest and highest values. Numbers in the legend refer to the number of trainable parameters.%
    \label{fig:phc}}
\end{figure}

We demonstrate the QNM-Net on two example systems. The first example is a \gls{phc} slab, consisting of a sheet of lossless dielectric with a repeating pattern of holes.
We study transmission and reflection at normal incidence.
The design is described by five parameters $(h,r_1,r_2,r_3,r_4)$ as illustrated in \textbf{Figure~\ref{fig:phc}(a)}.
Four-fold rotational symmetry is enforced to ensure polarization independence, resulting in a $S$ matrix of size $2 \times 2$.
The scattering spectrum in the frequency range of interest is dominated by a single Fano resonance on an otherwise smooth background, as exemplified in \textbf{Figure~\ref{fig:phc}(b)}.

Having defined the system, we now choose the implementation of the QNM-Net submodels based on our prior knowledge about its physics.
Consulting the diagram in Figure~\ref{fig:qnm-net-architecture}, the background model should take $\boldsymbol{\varphi}$ and $\omega$ as inputs and return the corresponding background matrix $C(\omega)$. 
Because the \gls{phc} is reciprocal, lossless, and mirror symmetric, $C$ satisfies $C^T = C$, $C^\dagger C = I$, and $C_{11} = C_{22}$ \cite{benzaouia_quasi-normal_2021}, so we set
\begin{equation}\label{eq:phc_C}
    C =
    \begin{pmatrix}
        i\sin\alpha & \cos\alpha\\
        \cos\alpha & i\sin\alpha
    \end{pmatrix}e^{i\alpha}.
\end{equation}
To allow for a design-dependent slow variation with frequency, we let $\alpha=\alpha(\omega)$ be a linear function of $\omega$ with values at \SI{170}{\tera\hertz} and \SI{210}{\tera\hertz} predicted from $\boldsymbol{\varphi}$ using a single-layer \gls{nn}.
Next, the mode models should predict the \gls{qnm} frequencies $\tilde{\omega}_m$ and port amplitudes $\mathbf{d}_m$ from $\boldsymbol{\varphi}$, with each mode model corresponding to one mode in the \gls{qnm} expansion.
Since the scattering spectrum of the \gls{phc} is well approximated by a single resonance, we only use one mode model in this case. 
Furthermore, because the \gls{phc} is reflection symmetric, the \glspl{qnm} are either even or odd, corresponding to $\mathbf{d}_m = (1,1)$ or $\mathbf{d}_m = (1,-1)$, respectively~\cite{alpeggiani_quasinormal-mode_2017, benzaouia_quasi-normal_2021}.
The observed Fano mode is even, so we set $\mathbf{d}_1 = (1,1)$.
The material of the \gls{phc} being lossless implies $\gamma_{\mathrm{nr},1}=0$.
We let the remaining unknown and design-dependent mode parameters $\omega_1$ and $\gamma_{\mathrm{r},1}$ be predicted from $\boldsymbol{\varphi}$ using a single-layer \gls{nn}, with causality enforced by applying a strictly positive activation function to $\gamma_{\mathrm{r},1}$~\cite{benzaouia_quasi-normal_2021}.
The embedding model should predict $\tau_n$ from $\boldsymbol{\varphi}$. 
We use a single-layer \gls{nn} to predict $\tau_1$ from $\boldsymbol{\varphi}$ and set $\tau_2=\tau_1$ to respect reflection symmetry.
Finally, we use a fully-connected feed-forward \gls{nn} in the feature extractor to provide the bulk of the trainable weights of the network.
A linear skip connection was added to improve training robustness and convergence speed~\cite{xu_physics-informed_2024}. A detailed schematic of all the submodels is provided in the Supporting Information.

To train the QNM-Net, we generated a dataset of \num{10000} random \gls{phc} designs and their corresponding scattering spectra $S(\omega)$ at 200 equally spaced frequency points using frequency-domain finite-element method simulations.
The QNM-Net was trained to minimize the $S$-parameter mean-squared-error ($S$-MSE), defined as
\begin{equation*}
    \textnormal{$S$-MSE} = \frac{1}{N_S N_\omega}\sum_{i,j,k} |S^\textnormal{predicted}_{ij}(\omega_k) - S^\textnormal{true}_{ij}(\omega_k)|^2,
\end{equation*}
where $N_S$ is the number of elements $S_{ij}$ in the scattering matrix and $N_\omega$ is the number of sampled frequencies $\omega_k$.
Gradients were automatically backpropagated through the entire model, including the \gls{qnm}-expansion, using automatic differentiation.
Models were trained on \SI{80}{\percent} of the dataset, while the remaining \SI{20}{\percent} was used for validation. The validation error was averaged over five dataset splits.
More details about the neural-network implementation and training procedures are provided in the Supporting Information.

The trained QNM-Net is able to reproduce the simulated scattering accurately, as exemplified by the sample shown in Figure~\ref{fig:phc}(b).
The sample has an $S$-MSE of $2.1\times10^{-4}$, which is close to the mean of the loss histogram shown in \textbf{Figure~\ref{fig:phc}(c)}.
The histogram was evaluated on a test set separate from both the validation and training sets to avoid bias.
For reference, we compare the \gls{qnm}-Net to three standard fully-connected feed-forward \glspl{nn} with varying numbers of trainable parameters, similar to architectures used in forward models in the literature~\cite{deng_neural-adjoint_2021, sarkar_physics-informed_2023, xu_enhanced_2020, so_simultaneous_2019, liu_training_2018}.
\textbf{Figure~\ref{fig:phc}(d)} shows the validation loss for all models at different dataset sizes.
The parameter count of all models are listed in the legend of the figure.
The QNM-Net requires only \SI{2}{\percent} of the dataset, corresponding to 160 training samples, to achieve an $S$-MSE less than $10^{-3}$.
The reference models require approximately one order of magnitude more data and trainable parameters to achieve a similar loss.
This shows that the QNM-Net can reduce the need for training data significantly by learning the physics of the system in terms of its \glspl{qnm}.

\begin{figure}[hbt!]
    \includegraphics{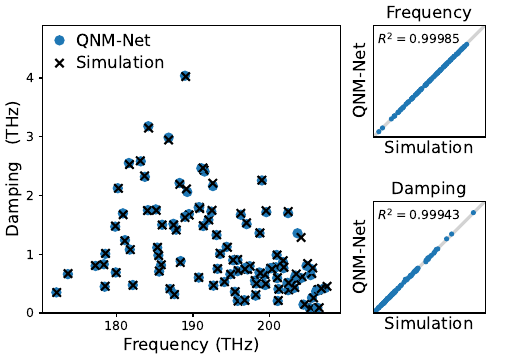}
    \caption{%
        Comparison between $\tilde{\omega}_1$ learned by the QNM-Net and calculated by full-wave eigenmode simulations for the PhC geometry.
        Left: QNM frequency and damping for 100 randomly generated designs. Each dot/cross corresponds to a separate design. Right: Parity plots for frequency and damping with corresponding coefficient of determination.
    \label{fig:compare_to_eigenmode}}
\end{figure}

Another benefit of physics-informed models operating according to the principle shown in the schematic in Equation~\eqref{eq:physics-informed-principle} is that their predictions are explainable in terms of the network-predicted physics parameters \cite{khatib_learning_2022}, which for the \gls{qnm}-Net are $C(\omega)$, $\tilde{\omega}_m$, $\mathbf{d}_m$, and $\tau_n$.
Based on the theoretical foundations of the \gls{qnm} expansion, we know that $\tilde{\omega}_m$ correspond exactly to eigenfrequencies of Maxwell's equations \eqref{eq:maxwell_equations}.
Thus, the accuracy of the learned physics can be verified by comparing the network-predicted $\tilde{\omega}_m$ to eigenfrequencies calculated using full-wave eigenmode simulations.
\textbf{Figure~\ref{fig:compare_to_eigenmode}} shows the network-predicted $\tilde{\omega}_1$ and the closest simulated eigenfrequency for 100 new randomly generated designs for the \gls{phc} slab example.
The two methods agree to high precision, indicating that the network training has converged to a representation of the spectra that matches the correct underlying physics for this system.
It should be noted that, although finding the correct \gls{qnm} parameters should lead to vanishing S-MSE and thus correspond to minima in the \gls{nn} loss landscape, there is no guarantee that these minima are unique or that the training procedure will always find them.
Thus, validating the trained model with a reference eigenmode solver as shown here is necessary if the accuracy of the learned physics is of importance.

\begin{figure}[hbt!]
    \includegraphics{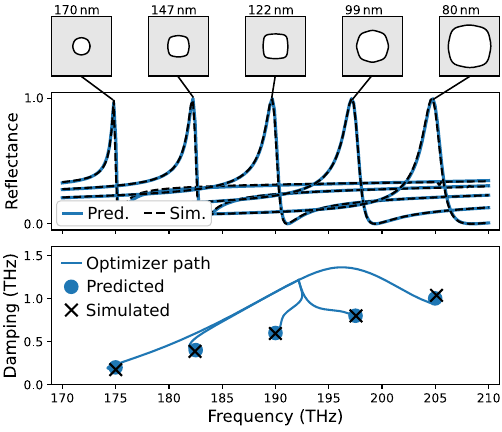}
    \caption{%
        Inverse design of \gls{phc} slabs.
        The \gls{phc} geometries in the top row were optimized for linearly spaced complex eigenfrequencies.
        The thickness of the PhC slab is displayed above each design.
        QNM-Net predictions (blue) and full-wave simulations (black) of the \gls{phc} reflectivity spectra and eigenmodes are shown in the middle and lower panels.
    \label{fig:inverse_design}}
\end{figure}

Since the QNM-Net is able to accurately predict the \gls{qnm} parameters of the PhC, these can now be used as design targets for inverse design.
To demonstrate this, we inverse design five \gls{phc} slabs with linearly increasing eigenfrequency and loss rate.
Starting from an initial design with all design parameters set to zero, the difference between model-predicted and desired value of $\tilde{\omega}_1$ was minimized using the Adam optimizer \cite{kingma_adam_2017} with gradients backpropagated through the network using automatic differentiation.
The optimization converged to the desired eigenfrequency after a few hundred steps in less than one second.
The final designs were simulated using the full-wave solver as verification.
The predicted and simulated results shown in \textbf{Figure~\ref{fig:inverse_design}} match closely, demonstrating that a \gls{qnm}-Net trained solely on spectral data can be used for fast and accurate inverse design of eigenfrequencies.

\subsection{Free-form dielectric metasurface}
\begin{figure}[hbt]
    \includegraphics{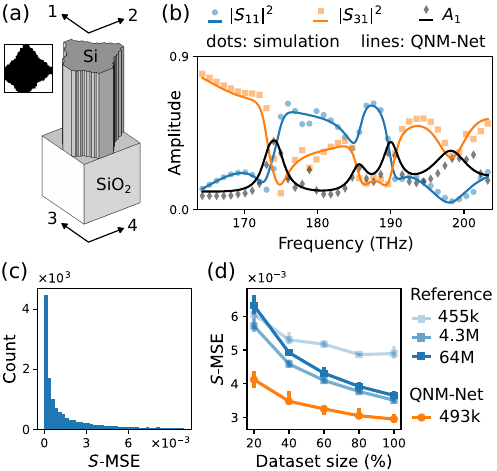}
    \caption{
        QNM-Net applied to free-form metasurface.
        (a)~Illustration of a metasurface unit cell and the corresponding bitmap. Numbered arrows indicate the locations and polarizations of the ports.
        (b)~Simulated and QNM-Net-predicted spectrum. $A_1=1-\Sigma_n |S_{n1}|^2$ is the absorbed power fraction for incidence on port 1.
        (c)~Histogram of QNM-Net test losses for dataset size \SI{80}{\percent}.
        (d)~Validation loss as a function of dataset size for the QNM-Net and reference models. Numbers in the legend refer to the number of trainable parameters.
    \label{fig:metasurface}}
\end{figure}

To investigate if the QNM-Net approach is also beneficial for more complex geometries when less is known about the physics, we apply the QNM-Net to free-form all-dielectric metasurfaces.
We use a dataset introduced in Reference~\cite{gahlmann_evaluation_2025} containing roughly \num{80000} meta-atoms and their corresponding scattering spectra 
(more details are available in the Supporting Information).
The geometry is described by a $100\times100$ binary bitmap representing the etching mask for the metasurface unit cell, as illustrated in \textbf{Figure~\ref{fig:metasurface}(a)}.
As far as we know, no previous works have applied \glspl{nn} with resonance-based physics models to free-form designs.
This system has a vastly larger design space, lower frequency resolution, a polarization-dependent scattering response, multiple overlapping resonances, nonradiative loss, and lacks mirror symmetry due to the presence of a substrate.
Furthermore, only two columns of the $S$ matrix, corresponding to light incident from the top, are available in the dataset.

We again adapt the QNM-Net submodel implementation to our prior knowledge about the system.
Because the design parameterization now has an inherent spatial structure, we use a version of the convolutional DenseNet architecture~\cite{huang_densely_2018} which we previously found to perform well on this dataset \cite{gahlmann_deep_2022} as the feature extractor.
In this case, there is no clearly distinguishable background. 
We therefore set $C=-I$, which is commonly used in \gls{qnm} expansions~\cite{alpeggiani_quasinormal-mode_2017, benzaouia_quasi-normal_2021} and let the QNM-Net fit the background using many broad resonances.
The spectra have a varying number of modes within the sampled frequency range.
We chose to include 20 mode models in the QNM-Net, which we deemed to be more than sufficient based on manual inspection of a few simulated spectra.
If a smaller number of modes are necessary to explain a given spectrum, the QNM-Net can learn to place the superfluous modes outside the sampled range, so that they do not contribute to the predicted spectrum.
For each mode, the QNM frequency and damping rates, as well as the real and imaginary parts of $\mathbf{d}_m$, are predicted from $\boldsymbol{\varphi}$ by small \glspl{nn}.
Each mode model has separate trainable weights.
Fixed port delays $\tau_n$ are used to account for the distance between the metasurface boundary and the ports.

Subsequently, we compare the QNM-Net to reference models consisting of DenseNets followed by branched feed-forward \glspl{nn}, similar to the architecture used in our previous work \cite{gahlmann_deep_2022}.
The same DenseNet architecture was used for all models, but the width of the fully-connected layers was varied to change the number of trainable parameters. \textbf{Figure~\ref{fig:metasurface}(b)} shows a simulated spectrum and the corresponding QNM-Net prediction for a sample from the validation set.
The $S$-MSE of the prediction is $2.4\times10^{-3}$, which is typical compared to the histogram shown in \textbf{Figure~\ref{fig:metasurface}(c)}.
We observe that the QNM-Net reproduces most resonances, but fails to predict a few weaker resonances. We believe this is due to the design space being so large that even tens of thousands of training samples are not enough to accurately learn all features of the spectra.
Nevertheless, \textbf{Figure~\ref{fig:metasurface}(d)} shows that the QNM-Net requires only around one third of the data to achieve a loss as low as that of the best reference models, demonstrating that the method is useful for spectrum prediction also for this system even if all predicted resonances are not accurate.

\begin{figure}[hbt]
    \includegraphics{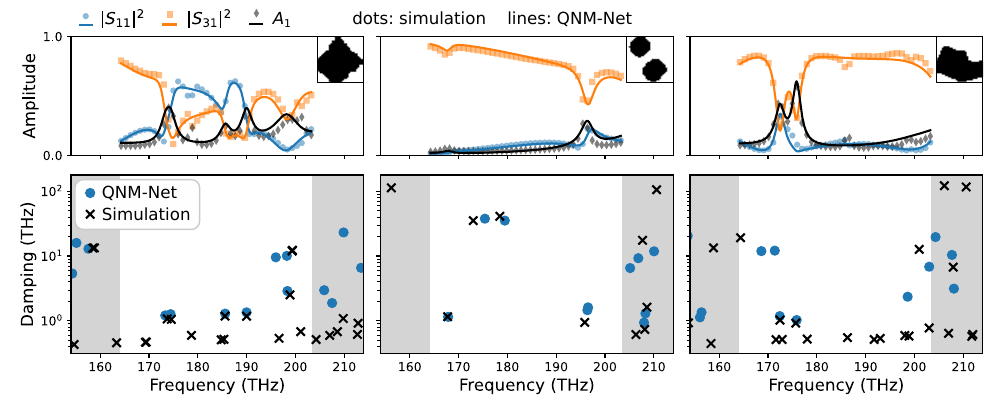}
    \caption{
    Eigenmode study for the metasurface example. Each column shows the QNM-Net-predicted and simulated spectrum (top row) and eigenfrequencies (bottom row) of a different design (top-right inset) from the validation set. Shaded regions in the bottom row correspond to frequencies outside the training range.
    \label{fig:metasurface_eigenmodes}}
\end{figure}

To further investigate the accuracy of the learned \glspl{qnm}, we conduct an eigenmode study for three metasurface designs from the validation set with varying number of resonances in the sampled frequency range.
The QNM-Net-predicted eigenfrequencies together with all eigenfrequencies returned by an eigenmode solver with outgoing radiation boundary conditions are shown in \textbf{Figure~\ref{fig:metasurface_eigenmodes}}.
Most modes predicted by the QNM-Net inside the sampled frequency range correspond to true eigenmodes found by the simulator.
Modes outside this range and high-damping modes close to the edges are generally not accurate, which is expected since there is very little information available in the data to learn these modes.
Interestingly, designs that connect the edges of the unit cell (the leftmost and rightmost designs in \textbf{Figure~\ref{fig:metasurface_eigenmodes}} exhibit several low-damping modes with no clear corresponding features in the scattering spectra and no corresponding modes predicted by the QNM-Net.
We interpret these as modes that are very weakly coupled to the incoming radiation and therefore do not contribute significantly to the scattering spectrum.
This illustrates that the QNM-Net automatically identifies the modes that are the most important for the scattering response, whereas eigenmode solvers return very large numbers of modes that would otherwise be cumbersome to filter manually.

\section{Conclusion and Further Research}
In conclusion, we have demonstrated a modular physics-informed \gls{nn} architecture for electromagnetic scattering based on the \gls{qnm} expansion of the scattering matrix.
We have shown that this architecture requires significantly less trainable parameters and is substantially more data efficient than conventional \glspl{nn} for prediction of scattering spectra.
Compared to previous physics-informed \glspl{nn} for electromagnetic scattering \cite{xu_physics-informed_2024, blanchard-dionne_teaching_2020, khatib_learning_2022}, we expect the QNM-Net to perform similarly in terms of accuracy and data efficiency while having the benefit of being applicable to a wider range of systems and a more well-founded theoretical basis in the formalism of quasinormal modes.
In particular, the \gls{qnm} frequencies learned by the model correspond directly to the eigenmodes of the scatterer, which was verified by comparing the learned physics to eigenmode simulations.
We believe that the method will be useful for modeling a wide range of electromagnetic systems where the scattering spectrum is accurately described by a set of resonances.

The method presented here opens the door to further developments in machine learning for electromagnetic scattering, which previously was limited to overly simplified systems or required an amount of training data that was difficult to obtain in practice.
The significantly improved data efficiency can make it practically feasible to train models on experimental data for simple designs.
In the Supporting Information, we add simulated noise to the training data to demonstrate that the QNM-Net is more robust to noise than standard \glspl{nn}, which indicates that the QNM-Net may have additional advantages in this scenario.
The QNM-Net can also be combined with methods such as those presented in References~\cite{raza_fabrication-aware_2025, jenkins_establishing_2021} to account for fabrication-induced performance degradation in inverse design.
Furthermore, investigating the learned parameters can provide insights into the underlying physics for automated knowledge discovery.
For example, if the reflection symmetry of the PhC mode had not been included as a prior constraint, it could have been discovered by looking at the model predictions for $\mathbf{d}_1$.

In future studies, it would be interesting to investigate the effect of individual constraints to determine which inductive biases are most important for data efficiency. 
Another possible development is to use transfer learning of the QNM-Net submodels for further improvements to data efficiency. For example, the feature extractor could be pretrained as the encoder network of an autoencoder.
The rigorous connection between the QNM expansion and electromagnetic eigenmodes also makes it possible to train the QNM-Net on eigenmode simulations.
This approach could be particularly useful for sharp resonances that are not easily resolved in the frequency domain.

\medskip
\textbf{Acknowledgements} \par
We acknowledge partial financial support from Chalmers’ area of advance Nano (Excellence Grant), from the Swedish Research Council under Grant No.~2020-05284, and from the Knut och Alice Wallenberg Stiftelse under Grant No.~2022.0090. Training data generation and NN training were performed on resources provided by the Swedish National Infrastructure for Computing (NAISS), at the Chalmers/C3SE and KTH/PDC sites, partially funded by the Swedish Research Council under Grant No.~2022-06725. The work was performed in part within the framework of the Excellence Center META-PIX.

\medskip
\textbf{Data Availability} \par
The photonic crystal dataset and code for implementing the QNM-Net are publicly available in the GitHub repository github.com/ViktorLilja/qnm-net.
The freeform metasurface dataset and code used to generate the results and figures are available from the authors upon request.

\medskip

\bibliographystyle{spiejour}
\bibliography{references}

\end{document}